\documentclass[prb,twocolumn,superscriptaddress,showpacs]{revtex4-2}

\usepackage{amsmath}    
\usepackage{dcolumn}   
\usepackage{graphicx}  
\usepackage{bm}        
\usepackage{amssymb}   

 \usepackage{color}
  \usepackage{ulem}
  \newcommand{\Nadd}[1]{#1}
 \newcommand{\Nrem}[1]{}

\newcommand{\tens}[1]{\underline{\underline{#1}}}
\newcommand{\bfr}{{\bf r}}

\begin{document}

\title{Constructing semilocal approximations for noncollinear spin density functional theory featuring exchange-correlation torques}

\author{Nicolas Tancogne-Dejean}
  \email{nicolas.tancogne-dejean@mpsd.mpg.de}
 \affiliation{Max Planck Institute for the Structure and Dynamics of Matter and Center for Free-Electron Laser Science, Luruper Chaussee 149, 22761 Hamburg, Germany}
 \affiliation{European Theoretical Spectroscopy Facility (ETSF)}

 \author{Angel Rubio}
\affiliation{Max Planck Institute for the Structure and Dynamics of Matter and Center for Free-Electron Laser Science, Luruper Chaussee 149, 22761 Hamburg, Germany}
 \affiliation{European Theoretical Spectroscopy Facility (ETSF)}
 \affiliation{Nano-Bio Spectroscopy Group, Universidad del Pa\'is Vasco UPV/EHU, 20018 San Sebasti\'an, Spain }
\affiliation{Center for Computational Quantum Physics (CCQ), The Flatiron Institute, 162 Fifth Avenue, New York, New York 10010, USA}

 \author{Carsten A. Ullrich}
 \affiliation{Department of Physics and Astronomy, University of Missouri, Columbia, Missouri 65211, USA}

\begin{abstract}

We present a semilocal exchange-correlation energy functional for noncollinear spin density functional theory based on short-range expansions
of the spin-resolved exchange hole and the two-body density matrix.
Our functional is explicitly derived for noncollinear magnetism, U(1) and SU(2) gauge invariant, and gives rise to nonvanishing exchange-correlation torques.
Testing the functional for frustrated antiferromagnetic chromium clusters, the exchange part is shown to perform favorably compared to the far more expensive
Slater and optimized effective potentials, and a delicate interplay between exchange and correlation torques is uncovered.
\end{abstract}

\maketitle

\section{Introduction}
In the framework of density functional theories, the key to success in describing the equilibrium and time-dependent properties of electronic systems
lies in the quality of approximations of the exchange-correlation (xc) energy and its functional derivatives.
In the quest to develop more accurate xc functionals, most of the recent effort has been devoted to the construction of approximations for spin unpolarized systems or for systems well described by collinear spins \cite{Perdew2013,Becke2014,Su2017,Teale2022}.

Considering the ever growing interest in the fields of spintronics and optical control of magnetism, reliable functionals going beyond the widely used local spin density approximation (LSDA) are highly desirable.
In particular, one of the features expected from the exact xc magnetic field, $\mathbf{B}_{\mathrm{xc}}$, of spin density functional
theory (SDFT) \cite{Barth1972,Gunnarsson1976} is to generate locally a non-zero xc torque, which arises from the fact that in SDFT the Kohn-Sham current is not the same as the exact many-body current \cite{PhysRevLett.87.206403}. Globally, this xc torque must average to zero, which is known as the zero-torque theorem \cite{PhysRevLett.87.206403}.
However, the practical significance of xc torques is still relatively little explored \cite{Bulik2013,Pluhar2019}, and a better understanding would be beneficial for the
application of SDFT to noncollinear magnetism and spin dynamics \cite{Krieger2015}.

In frustrated antiferromagnetic systems, such as free-standing Cr monolayers, local xc torques can be sizable in the vicinity of the Cr atoms~\cite{PhysRevLett.98.196405}, based on calculations performed at the level of the optimized effective potential (OEP) for exact exchange (EXX) \cite{Krieger1992,Kummel2008}.
How this nonzero xc torque affects the static and dynamical properties of magnetic systems, such as light-induced demagnetization \cite{Krieger2015} or the dispersion of spin waves \cite{NTD2020},
is an open question that needs to be addressed in order to gain understanding and control of the magnetization dynamics in these systems.
More recently, it was demonstrated using a source-free version of the LSDA \cite{sharma2018source} that the xc torque affects the light-demagnetization process \cite{dewhurst2018effect}.
\Nadd{The influence of xc torques on magnetization dynamics was also recently studied in small Hubbard clusters\cite{2301.01509}.}

Two main approaches are available for constructing xc functionals for noncollinear systems. The most common approach consists in promoting existing functionals for collinear systems to work for noncollinear ones. As shown by K\"ubler and others \cite{kubler1988density,Sandratskii1998}, noncollinear LSDA calculations
can be carried out by performing a rotation of the spin density matrix in the frame of the local magnetization $\bf m$. The drawback of this is clear: because the resulting local $\mathbf{B}_{\mathrm{xc}}$ is aligned with $\bf m$, it cannot produce any local xc torques. Various extensions of this approach to generalized spin gradient approximation (GGA) functionals have been proposed since then \cite{PhysRevB.75.125119,scalmani2012new}. While these different schemes can produce non-zero xc torques, they tend to suffer from numerical instabilities and can encounter difficulties in reproducing the collinear limit unless special care is taken \cite{Desmarais2021}. \Nadd{We also mention a recently proposed multicollinear averaging scheme~\cite{PhysRevResearch.5.013036}.}

In this work, we follow a second approach: instead of upgrading existing collinear xc functionals, 
we directly construct xc functionals for noncollinear magnetism. This philosophy was used in some earlier studies, such as the transverse spin-gradient functional of Eich and Gross \cite{PhysRevLett.111.156401}.
More recently, Pittalis and coworkers \cite{PhysRevB.96.035141}
proposed a general method to construct noncollinear-spin density functionals from existing collinear-spin functionals by applying a simple minimal substitution on the kinetic energy density and the on-top hole expression, which controls the short-range behavior of the exchange hole.
They demonstrated the method by constructing the spin non-collinear version  of the Becke-Roussel meta-GGA (MGGA) exchange energy functional~\cite{PhysRevA.39.3761}, but no numerical implementation of this functional has been reported so far. This is the functional we use here as noncollinear exchange functional.
Finally, a class of orbital-dependent functionals for noncollinear SDFT was proposed in Ref. \onlinecite{PhysRevB.98.035140} and applied to the asymmetric Hubbard dimer and other small lattice systems \cite{Pluhar2019}.

In this \Nrem{Letter}\Nadd{paper}, we show how to use a short-range expansion of the exchange hole and of the two-body reduced density matrix to build an exchange-correlation MGGA functional for noncollinear SDFT. Our derivation of the exchange part differs slightly from that of  Ref.~\onlinecite{PhysRevB.96.035141} but leads to the same result.
A matching correlation functional will be obtained via a generalization of the correlated wavefunction ansatz proposed originally by Colle and Salvetti~\cite{colle1975approximate}, which allows us to generalize the work of Lee, Yang, ad Parr (LYP)~\cite{PhysRevB.37.785} to the noncollinear case.
From this, we obtain an MGGA xc functional that not only depends locally on the direction of $\bf m$ but also on the direction of the curvature of the exchange hole, which implies that the resulting $\mathbf{B}_{\mathrm{xc}}$ can exert a local torque on $\bf m$.
We then apply this functional to planar Cr clusters with frustrated antiferromagnetic interactions and show that not only the functional produces a nonzero exchange torque, but that it properly reproduces the most salient features of the exchange torques obtained by the far more complicated noncollinear Slater potential and EXX-OEP.
We mention that it was shown for model systems that Slater and EXX-OEP yield decent approximations to the exact xc torques, as long as the systems are not too strongly interacting \cite{Pluhar2019}.

\section{Noncollinear exchange functional}

The starting point for the exchange part is the exchange energy of a system of $N$ electrons,
\begin{equation}
 E_{\mathrm{x}} = - \frac{1}{2} \int \int \frac{d\mathbf{r}d\mathbf{r'}}{|\mathbf{r}-\mathbf{r'}|} \mathrm{Tr}\Big[\tens{\gamma}(\mathbf{r},\mathbf{r'}) \tens{\gamma}(\mathbf{r'},\mathbf{r}) \Big] \,,
 \label{eq:energy_exchange_hole}
\end{equation}
where $\mathrm{Tr}$ is the trace over spin indices of the one-particle spin density matrix
$\gamma_{\sigma\tau}(\mathbf{r},\mathbf{r'}) = \sum_j^N \psi_{j\sigma}(\mathbf{r})\psi^*_{j\tau}(\mathbf{r'})$,
constructed from two-component
spinor Kohn-Sham orbitals, where $\sigma = \uparrow,\downarrow$ and likewise for $\tau$.
Here and in the following, doubly underlined quantities such as $\tens{\gamma}$ represent
$2\times 2$ matrices in spin space.
Here, $E_{\rm x}$ is U(1)$\times$ SU(2) gauge invariant \cite{PhysRevB.96.035141}; directly
approximating $E_{\rm x}$ is therefore a good strategy to produce meaningful exchange functionals for noncollinear SDFT.

From $E_{\rm x}$ of Eq. (\ref{eq:energy_exchange_hole}) the noncollinear OEP EXX potential as well as the Slater potential can be derived
for noncollinear spins \cite{PhysRevB.98.035140}; it also generalizes the definition of the exchange hole,
as discussed in \Nrem{the Supplementary Material (SM) \cite{supp}}\Nadd{App.~\ref{app:exchange_hole}}.
Since the noncollinear exchange hole is now a $2\times 2$ matrix in
spin space, it would be necessary to approximate its diagonal as well as off-diagonal elements. The existing models for the collinear exchange hole \cite{doi:10.1063/5.0031995} rely on the fact that the modeled exchange is positive and normalized to unity. These properties are obviously not fulfilled by the off-diagonal terms of the noncollinear exchange hole, which makes them difficult to approximate using existing collinear models.

To circumvent this problem, one can seek an alternative quantity to approximate, for which we could use already existing collinear models.
We rewrite Eq. (\ref{eq:energy_exchange_hole}) as
\begin{equation}
 E_{\mathrm{x}} = - \frac{1}{2}  \int d\mathbf{r}n(\mathbf{r}) \int d\mathbf{r'} \frac{h_{\rm x}(\mathbf{r},\mathbf{r'})}{|\mathbf{r}-\mathbf{r'}|}  \,,
 \label{eq:energy_eff}
\end{equation}
where $n(\mathbf{r}) = \sum_\sigma n_{\sigma\sigma}(\mathbf{r})$ is the total density, and the quantity
$h_{\rm x}(\mathbf{r},\mathbf{r'}) = \mathrm{Tr}\big[\tens{\gamma}(\mathbf{r},\mathbf{r'}) \tens{\gamma}(\mathbf{r'},\mathbf{r})\big]/n(\mathbf{r})$,
referred to in the following as the effective exchange hole, displays the same properties as the physical exchange hole for unpolarized or collinear-spin systems. Indeed, $\mathrm{Tr}\big[\tens{\gamma}(\mathbf{r},\mathbf{r'}) \tens{\gamma}(\mathbf{r'},\mathbf{r})\big]$ is always positive, and hence $h_{\rm x}$ is always positive, too. Moreover, $h_{\rm x}$ has the right normalization condition,  $\int d\mathbf{r'}  h_{\rm x}(\mathbf{r},\mathbf{r'}) = 1$. Therefore,
$h_{\rm x}$ is suited to be approximated by already existing models for the collinear exchange hole. A recent review
\cite{doi:10.1063/5.0031995} of the existing models for the collinear exchange hole found that the hydrogenic model \cite{PhysRevA.39.3761} seems to perform best, at least for the systems considered.

Only the spherical average of the effective exchange-hole function around a given reference point $\mathbf{r}$ is relevant for the exchange energy.
\Nrem{As shown in the SM~\cite{supp}, performing the Taylor expansion of the spherical average of $h_{\rm x}$ to obtain its short-range behavior around a reference point $\mathbf{r}$ gives an expression of $h_{\rm x}(s)$ (where $s$ is the radius centered about $\bfr$)
in terms of an effective kinetic energy density $\tilde{\tau}$, whose expression is given in the SM \cite{supp}.}
\Nadd{
We perform a Taylor expansion of the spherical average of $h_{\rm x}$ to obtain its short-range behavior around a reference point $\mathbf{r}$, which gives, up to second order in the distance $s$ from $\mathbf{r}$, its on-top value and its curvature.
Omitting the explicit dependence on $\mathbf{r}$, we obtain
\begin{eqnarray}
h_{\rm x}(s) &=& \sum_\sigma \frac{n_{\sigma\sigma}^2
+ |n_{\sigma-\sigma}|^2 }{n} + \frac{s^2}{6n} \sum_\sigma \Big[n_{\sigma\sigma}\nabla^2n_{\sigma\sigma}
\nonumber\\
&&+  \Re(n_{\sigma-\sigma}\nabla^2n_{-\sigma\sigma})  - 2n_{\sigma\sigma}\tilde{\tau}_{\sigma\sigma} -2 \Re(n_{\sigma-\sigma}\tilde{\tau}_{-\sigma\sigma})
\nonumber\\
&&+\frac{1}{2}(|\nabla n_{\sigma\sigma}|^2  + |\nabla n_{\sigma-\sigma}|^2) \Big]  + {\cal O}(s^4).
\end{eqnarray}
Here, $\tilde{\tau}$ is an effective kinetic energy density defined by the matrix equation
\begin{equation}
\tens{n}\,\tens{\tilde{\tau}} + \tens{\tilde{\tau}}\,\tens{n} = \tens{n}\,\tens{\tau} + \tens{\tau}\,\tens{n}  -2i\nabla(\tens{n}\,\tens{j}-\tens{j}\,\tens{n})- 2\tens{\mathbf{j}}\cdot\tens{\mathbf{j}} \:,
\label{eq:tau_U1}
\end{equation}
with $\tau_{\alpha\beta} = \sum_j (\nabla\psi_{j\alpha}(\mathbf{r}))\cdot \nabla\psi_{j \beta}^*(\mathbf{r})$ the usual kinetic energy density, and $\tens{\mathbf{j}}$ is the paramagnetic current density.}

This quantity, $\tilde{\tau}$, is U(1) gauge invariant but is not invariant under a local SU(2) gauge transformation~\cite{PhysRevB.96.035141}. Of course, the entire curvature is SU(2) invariant.
Thus, in order to obtain an expression for $h_{\rm x}(s)$ that is made of SU(2) invariant building blocks, we rewrite it as
\begin{eqnarray}\label{hxs}
\lefteqn{h_{\rm x}(s)=
n\zeta_x
+ \frac{s^2}{6} \left[\nabla^2n -2\gamma\left(\bar{\tau}-\tau^W\right) \right]} \\
&=&
n\zeta_x\bigg[1 + s^2 k_\mathrm{F}^2
\left(\frac{ 2}{3} q(n,|\mathbf{m}|) -\frac{\gamma}{5} \alpha(n, |\mathbf{m}|, |\nabla n|, \bar{\tau}) \right)\bigg],  \nonumber
\end{eqnarray}
where we define the dimensionless spin-polarization factor $\zeta_x(\mathbf{r}) = \frac{1}{2}(1 + \frac{|\mathbf{m}(\mathbf{r})|^2}{n(\mathbf{r})^2})$ which is obviously invariant under SU(2) local rotation; $\bar{\tau}$ is the U(1)$\times$SU(2) gauge-invariant kinetic energy density of Ref.~\onlinecite{PhysRevB.96.035141},  given by
\begin{eqnarray}
2n\bar{\tau} &=& \mathrm{Tr}\big[\,\tens{n}\,\tens{\tilde{\tau}} + \tens{\tilde{\tau}}\,\tens{n}\,\big]
+\sum_\sigma\bigg[ n_{\sigma\sigma}\nabla^2n_{\bar\sigma \bar\sigma} \nonumber\\
&-&\Re(n_{\sigma \bar\sigma}\nabla^2n_{\bar\sigma\sigma}) \left.
-\frac{|\nabla n_{\sigma \bar\sigma}|^2}{2}  +\frac{\nabla n_{\sigma\sigma}\cdot\nabla n_{\bar\sigma\bar\sigma}}{2}\right]
\end{eqnarray}
where $\bar\sigma = -\sigma$.
One can show (see \Nrem{SM \cite{supp}}\Nadd{App.~\ref{app:one_electron}}) that in the one-electron limit $\bar{\tau}$ reduces  to the von Weizs\"acker kinetic-energy density \cite{weizsacker1935theorie}
$\tau^{\rm W}=|\nabla n|^2/4n$ and in the uniform gas limit $\bar{\tau}$ reduces to $\bar{\tau}^{\mathrm{unif}} = \frac{3}{5} k_\mathrm{F}^2 n\zeta_x$,
with $k_\mathrm{F} = (6\pi^2n)^{1/3}$. Similar to common collinear MGGAs \cite{PhysRevLett.91.146401,PhysRevLett.115.036402}, we introduce a dimensionless parameter
$\alpha(n, |\mathbf{m}|, |\nabla n|, \bar{\tau}) = (\bar{\tau}-\tau^{\rm W})/\bar{\tau}^{\mathrm{unif}}$, which in the one-electron and uniform gas limits reduces to
$\alpha = 0$ and $\alpha = 1$, respectively, as in the collinear case.
We also define a dimensionless Laplacian $q(n,|\mathbf{m}|) = \nabla^2n/(4k_\mathrm{F}^2 n \zeta_x)$.
The different components of the exchange-hole curvature, in particular $\bar{\tau}$, are all U(1) and SU(2) gauge invariant.
Note that the SU(2) gauge invariance implies $\alpha \ge0$.

Let us comment here on the significance of the gauge invariance. Already at the level of collinear MGGAs, the U(1) gauge invariance is very important, especially for the dynamical case in which the current is non-zero or for the description of current-carrying states
\cite{PhysRevB.71.205107,PhysRevLett.95.196403,PhysRevA.80.032515,rasanen2010universal}. For noncollinear systems, a semi-relativistic theory including on equal footing electromagnetic fields and spin-related terms in the Hamiltonian (Zeeman term, spin-orbit coupling) should also preserve the local SU(2) gauge invariance~\cite{PhysRevB.96.035141}. This is the case of the exact-exchange energy, but also in our proposed functional. This is also needed if one wants to employ the generalized Bloch theorem~\cite{sandratskii1986energy}, for instance to study spin waves~\cite{NTD2020}.

It is clear that building a functional from the quantities ($\alpha$, $q$, $\bar{\tau}$) will directly recover the collinear limit, which is a strong requirement for any noncollinear xc functional. Moreover, we see that while the on-top term is determined by the direction of $\bf m$, the curvature has its own direction, which is independent of that of $\bf m$. This implies that the resulting $\mathbf{B}_{\mathrm{xc}}$ is in general not aligned with $\bf m$,
thus producing a nonvanishing local exchange torque. Finally, note that compared to Ref.~\onlinecite{PhysRevB.96.035141}, we introduced a scaling factor $\gamma$ in Eq. (\ref{hxs}) to ensure the correct
homogeneous electron gas limit, as done in the collinear case \cite{PhysRevA.39.3761}. This scaling factor does not break the gauge invariance of the energy, as it acts on a building block which is gauge invariant by itself. The limit of hydrogenic systems ($\alpha=0$) is also not affected by the choice of $\gamma$. In the following, we use $\gamma=0.8$ unless stated otherwise.

Having at hand the short-range behavior of the effective exchange hole, we now express it via a hydrogenic model \cite{PhysRevA.39.3761}, see \Nrem{SM \cite{supp}}\Nadd{App.~\ref{app:hydrogenic}}. Once the parameters $a(\mathbf{r})$ and $b(\mathbf{r})$ of the model are determined for each $\bf r$, we obtain the following expression for the exchange energy:
\begin{equation}\label{Ex}
E_{\mathrm{x}} = -3\frac{(3\pi^2)^{1/3}}{4\pi} \int d\mathbf{r} n^{4/3}(\mathbf{r})\zeta_x(\mathbf{r})^{1/3}  F_{\rm x}(\mathbf{r})\,.
\end{equation}
Here,
\begin{equation}
F_{\rm x}(\mathbf{r}) = \frac{4\pi^{2/3}e^{x(\mathbf{r})/3}}{3^{4/3}x(\mathbf{r}) }
\left[1-e^{-x(\mathbf{r})}\left(1+\frac{x(\mathbf{r}) }{2}\right) \right]
\end{equation}
plays the role of an enhancement factor,
where $x(\mathbf{r}) = a(\mathbf{r})b(\mathbf{r})\ge 0$.
Importantly, because the on-top effective exchange hole as well as the exchange hole curvature are U(1) and  SU(2) gauge invariant, $a$ and $b$ are invariant under local spin rotation, and our functional is then U(1) and SU(2) gauge invariant and also recovers naturally the collinear limit. Moreover, because the functional is based on the
exchange hole  of a physical system (the H atom), the energy is constrained to reasonable values.

\section{Noncollinear correlation functional}
Our correlation energy functional is obtained by extending the work of LYP~\cite{PhysRevB.37.785} to the noncollinear case. We start with the correlation energy
expression of Colle and Salvetti \cite{colle1975approximate},
\begin{equation}
 E_{\rm c} = -4a \!\! \int\!  d\mathbf{r} \frac{\rho_2(\mathbf{r},\mathbf{r})}{n(\mathbf{r})}\left[\frac{1+br_s^8(\bfr)[\nabla^2_\mathbf{s}\rho_2(\mathbf{r},\mathbf{s})]_{\mathbf{s=0}}e^{-c r_s(\bfr)}}{1+d r_s(\bfr)} \right]
\end{equation}
where $r_s(\bfr) = (4\pi n(\bfr)/3)^{-1/3}$ is the local Wigner-Seitz radius, $a=0.04918$, $b=0.132 (4\pi/3)^{8/3}$, $c=0.2533(4\pi/3)^{1/3}$, and $d=0.349(4\pi/3)^{1/3}$.
We then approximate the two-body reduced density matrix in the inter-particle coordinates as
\begin{equation}
 \rho_2(\mathbf{r}_1,\mathbf{r}_2) = \frac{1}{2}\Big[n(\mathbf{r}_1)n(\mathbf{r}_2) - \mathrm{Tr}\big[\tens{\gamma}(\mathbf{r}_1,\mathbf{r}_2) \tens{\gamma}(\mathbf{r}_2,\mathbf{r}_1)\big]\Big]\,.
\end{equation}

\Nadd{
 In order to evaluate the correlation energy, we need to evaluate $\rho_2(\mathbf{r},\mathbf{r})$ and  $[\nabla^2_\mathbf{s}\rho_2(\mathbf{r},\mathbf{s})]_{\mathbf{s=0}}$ with
\begin{equation}
 \rho_2(\mathbf{r}_1,\mathbf{r}_2) = \frac{1}{2}\Big[n(\mathbf{r}_1)n(\mathbf{r}_2) - \sum_{\alpha\beta}\gamma_{\alpha\beta}(\mathbf{r}_1,\mathbf{r}_2)\gamma_{\beta\alpha}(\mathbf{r}_2,\mathbf{r}_1)\Big]\,.
\end{equation}
We directly get from this definition the on-top value
\begin{equation}
 \rho_2(\mathbf{r},\mathbf{r}) = \frac{1}{2}n(\mathbf{r})^2 - \frac{1}{2}\mathrm{Tr}(\tens{n}\,\tens{n})
 = \frac{1}{4}n(\mathbf{r})^2 \Big(1- \frac{|\mathbf{m}(\mathbf{r})|^2}{n(\mathbf{r})^2}\Big) \,,
\end{equation}
where we define a spin-polarization factor $\zeta_c(\mathbf{r}) = (1- \frac{|\mathbf{m}(\mathbf{r})|^2}{n(\mathbf{r})^2})$ as introduced in the exchange functional.
After some algebra, the value of  $[\nabla^2_\mathbf{s}\rho_2(\mathbf{r},\mathbf{s})]_{\mathbf{s=0}}$ is found to be
\begin{eqnarray}
 [\nabla^2_\mathbf{s}\rho_2]_{\mathbf{s=0}} &=& -n(\tau^W-\frac{\nabla ^2 n}{4}) \nonumber\\
&+&  \frac{1}{4}\Big[ - \frac{1}{2}\mathrm{Tr}\big[\tens{n}\nabla^2\tens{n} + \nabla^2\tens{n}\tens{n}\big]\nonumber\\
&&
{} + 2\mathrm{Tr}\big[\tens{n}\,\tens{\tau} + \tens{\tau}\,\tens{n} \big] -4 \mathrm{Tr}\big[\tens{\mathbf{j}}\cdot\tens{\mathbf{j}}\big] \Big]\,,
 \end{eqnarray}
where we recognize the trace of the U(1) kinetic energy density $\tilde{\tau}$ defined by Eq.~\ref{eq:tau_U1}.\\
From these two expressions, we are getting the expression of the kinetic energy density.
\begin{equation}
 E_{\rm c} = -a \!\! \int\!  d\mathbf{r} n \zeta_c\left[\frac{1+br_s^5[\tau^{\rm HF} - \tau^W + (\nabla^2 n)/4]e^{-c r_s}}{1+d r_s} \right]\,,
\end{equation}
where 
\begin{equation}
 2n\tau^{\rm HF} =  \mathrm{Tr}\big[\tens{n}\,\tens{\tilde{\tau}}+ \tens{\tilde{\tau}}\,\tens{n}\big] - \frac{1}{4}\mathrm{Tr}\big[\tens{n}\nabla^2\tens{n} + \nabla^2\tens{n}\tens{n}\big]\,.
\end{equation}
This term is not directly connected to the gauge-invariant kinetic energy density introduced before. Moreover, it is not clear at this stage that the quantity $\left[ \tau^{\rm HF}-\tau^{\rm W} +\frac{\nabla^2 n}{4}\right]$ is SU(2) gauge invariant. It is clearly U(1) gauge invariant as $\tau^{\rm HF}$ is defined in terms of $\tens{\tilde{\tau}}$. In order to show that it is in fact gauge invariant, we first write $\tau^{\rm HF}$ in terms of the SU(2) gauge invariant kinetic energy $\bar{\tau}$
\begin{eqnarray}
\tau^{\rm HF}
&=& \frac{\bar{\tau}}{2}
 - \frac{1}{4n}\sum_\sigma\bigg[ n_{\sigma\sigma}\nabla^2n_{\sigma\sigma}
\nonumber\\
&+&
  n_{\sigma\sigma}\nabla^2n_{-\sigma-\sigma}
 -|\nabla n_{\sigma-\sigma}|^2
 +\nabla n_{\sigma\sigma}\cdot\nabla n_{-\sigma-\sigma}
 \bigg]\,. 
\end{eqnarray}
Now, using the fact that 
\begin{displaymath}
n\nabla^2 n = \sum_{\sigma} \big[ n_{\sigma\sigma}\nabla^2n_{\sigma\sigma} + n_{\sigma\sigma}\nabla^2n_{-\sigma-\sigma}\Big] ,
\end{displaymath}
we obtain that
\begin{eqnarray}
 \tau^{\rm HF}+\frac{\nabla^2 n}{4}
 = \frac{\bar{\tau}}{2}
 + \frac{1}{4n}\sum_\sigma\bigg[
 |\nabla n_{\sigma-\sigma}|^2
 -\nabla n_{\sigma\sigma}\cdot\nabla n_{-\sigma-\sigma}
 \bigg]\,.
\end{eqnarray}
In order to understand the meaning of the last term, we re-express it in terms of the components of the magnetization vector. Using that $n_{\uparrow\downarrow} = m_x+im_y$, we obtain easily that
\begin{equation}
 \sum_\sigma|\nabla n_{\sigma-\sigma}|^2 = \frac{1}{2}\big(  |\nabla m_x|^2 +  |\nabla m_y|^2\big)\,.
\end{equation}
Similarly, we find
\begin{equation}
 \sum_\sigma \nabla n_{\sigma\sigma}\cdot\nabla n_{-\sigma-\sigma} = \frac{1}{2}\big(  |\nabla n|^2 -  |\nabla m_z|^2\big)\,.
\end{equation}
Hence, we get that
\begin{eqnarray}
 \tau^{\rm HF}+\frac{\nabla^2 n}{4}
&=& \frac{\bar{\tau}}{2}
 + \frac{|\nabla \mathbf{m}|^2}{8n} - \frac{|\nabla n|^2}{8n}\nonumber\\
&=& \frac{\bar{\tau}}{2}
 + \frac{|\nabla \mathbf{m}|^2}{8n} - \frac{\tau^W}{2}
 \,.
\end{eqnarray}
This therefore leads to the correlation energy}
\Nrem{which yields after some algebra (see SM \cite{supp})}
\begin{equation}
 E_{\rm c} = -a \int d\mathbf{r} n \zeta_c \Big[ \frac{1+\frac{b r_s^{5}}{2}\big[ \bar{\tau}
 + \frac{|\nabla \mathbf{m}|^2}{4n} - 3\tau^W\big]e^{-c r_s}}{1+dr_s} \Big],
\end{equation}
where $\zeta_c(\mathbf{r}) = (1 - \frac{|\mathbf{m}(\mathbf{r})|^2}{n(\mathbf{r})^2})$.
It is straightforward to see that our noncollinear correlation functional is also
gauge invariant. Moreover, we are able to express it in terms of
the same gauge-invariant kinetic energy density $\bar{\tau}$ as in the noncollinear exchange functional of Eq.~\eqref{Ex}.

We note that in order to obtain a GGA correlation functional, as originally done by LYP, we would need to use here a gradient expansion up to second order of the noncollinear kinetic energy density. However, this gradient expansion for the SU(2) gauge invariant kinetic energy density is not known, except for the zero-order term which is $\bar{\tau}^{\mathrm{unif}}$. How the gradient expansion can be performed to maintain gauge invariance under local rotation of the spins needs to be carefully explored and we postpone this to a subsequent work. We therefore keep our correlation functional at the MGGA level.

\section{Results}
By making  the  total  energy  stationary  with  respect to spinor orbital variations, we obtain a differential operator rather than a local multiplicative Kohn-Sham potential, because of the explicit dependence on the kinetic energy density, see \Nrem{SM \cite{supp}}\Nadd{App.~\ref{app:potential}}.
This treatment, first used by Neumann, Nobes, and Handy \cite{neumann1996exchange}, is usually referred as generalized Kohn-Sham (gKS). We refer to MGGA treated using this approach as MGGA-gKS.
Alternatively, one can construct a local multiplicative Kohn-Sham potential using the OEP formalism (MGGA-OEP). This approach tends to be less used as the solution of the OEP equation can be numerically involved \cite{Kummel2008}. While these two approaches usually give similar total energies, they can yield different results for other quantities of interest such as the nuclear shielding of small molecules \cite{arbuznikov2003self} or the bandgap of solids \cite{PhysRevB.93.205205}, and it is therefore interesting to explore how gKS and OEP can differ for noncollinear systems.
In order to perform this analysis, we derived an explicit solution for the Krieger-Li-Iafrate (KLI) approximation \cite{Krieger1992} toward the exact OEP result for noncollinear spins (see \Nrem{SM \cite{supp}}\Nadd{App.~\ref{app:KLI}}). The implementation of this solution and of our MGGA functional was done using the real-space code Octopus~\cite{Octopus_paper_2019}.

As a critical check of the functional, we consider a planar Cr$_3$ cluster with frustrated antiferromagnetic interactions, which is typically used to test noncollinear versions of collinear functionals~\cite{PhysRevB.75.125119,scalmani2012new,PhysRevB.62.11556}.
Calculations were performed using a grid spacing of 0.15 Bohr, employing norm-conserving fully relativistic Hartwigsen-Goedecker-Hutter (HGH) pseudo-potentials~\cite{PhysRevB.58.3641}, including semicore electrons as valence ones and spin-orbit coupling in all the simulations. The distance between the Cr atoms is taken to be 3.7 Bohr.

\begin{figure}
  \begin{center}
    \includegraphics[width=\columnwidth]{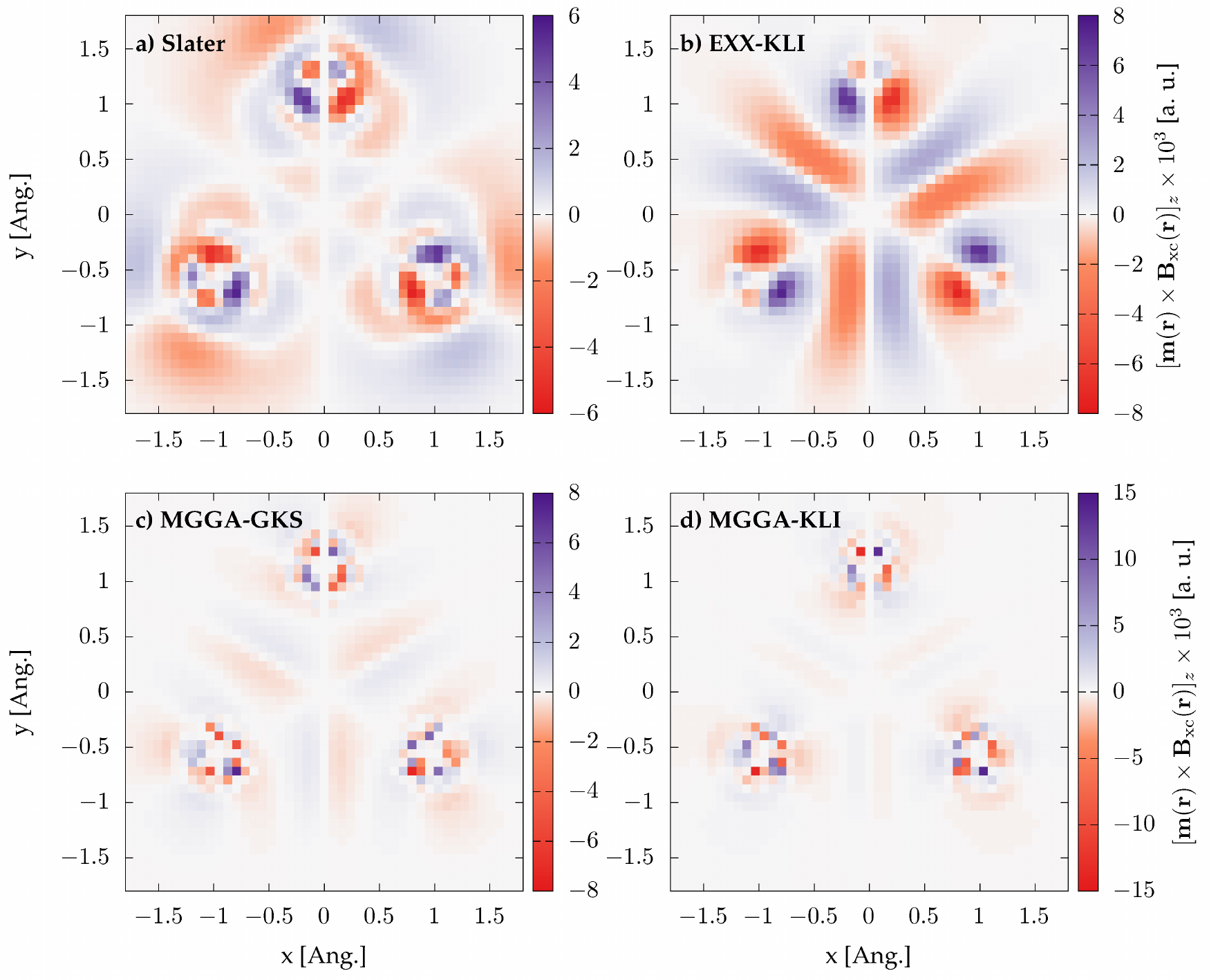}
    \caption{\label{fig:torque} The $z$ component of the local exchange torque $\mathbf{m}(\mathbf{r})\times\mathbf{B}_{\mathrm{x}}(\mathbf{r})$ around the Cr atoms in a
     Cr$_3$ cluster, computed from a) Slater potential, b) EXX-KLI, c) MGGAx functional at the gKS level, d) MGGAx at the OEP-KLI level.}
  \end{center}
\end{figure}

As an important measure of the performance of the functional, we first consider the local magnetic moment of the Cr atoms, computed on atom-centered spheres of radii 1.8 Bohr, see Table \ref{tab:table}. Interestingly, the exchange-only LSDA (LSDAx) gives much larger magnetic moments than the full LSDA,
showing the importance of correlation effects in this system.

The proposed MGGA exchange (MGGAx) functional significantly improves the description of the magnetic structure of the Cr$_3$ cluster compared to the LSDAx; all results are relatively insensitive to the choice of $\gamma$.
We note that there is a difference between using the MGGAx functional within the gKS framework and using the KLI scheme. This observation is in line with previously reported results for magnetic moments in solids \cite{PhysRevB.84.035117,PhysRevB.102.024407}, in which it was found, using the gKS approach, that MGGA for collinear magnetism only slightly changes the magnetic moments compared to the ones obtained by GGAs. Our results for Cr$_3$ reveal that more detailed studies of the impact on the choice of gKS versus OEP for treating MGGA in magnetic systems is needed.
We note that the LSDA correlation reduces the magnetic moment compared to a LSDAx calculation. A similar effect is obtained for our correlation functional, which reduces the magnetic moment compared to exchange-only.
We also report in Table \ref{tab:table} the ionization potential computed from the different functionals. Similar conclusions are obtained as from the magnetic moments. In particular, LSDA correlations and our MGGA correlations increase the ionization potential compared to exchange-only.

\begin{table}
\caption{\label{tab:table} Local magnetic moment $|\mathbf{m}|$, in $\mu_B$, and ionization potential $I_p$, in eV,
of the Cr atoms in Cr$_3$ obtained for different levels of theory (see text).
}
\begin{ruledtabular}
\begin{tabular}{lcc}
Functional & $|\mathbf{m}|$ & $I_p$\\
\hline
LSDA & 1.67 & 2.90 \\
LSDAx &  2.66 & 2.30 \\
LSDAx+MGGAc-gKS &  1.81  & 2.60 \\
MGGAx+MGGAc-gKS ($\gamma=0.8$) &  2.30  & 4.61 \\
MGGAx-gKS ($\gamma=0.8$)& 3.04 & 3.65\\
MGGAx-gKS ($\gamma=1$) & 3.07 & 3.53\\
MGGAx-KLI ($\gamma=0.8$)& 3.09 & 3.59\\
MGGAx-KLI ($\gamma=1$)& 3.14 & 3.47\\
Slater & 3.48 & 6.52\\
EXX-KLI & 3.81 & 4.68\\
Hartree-Fock & 3.86 & 4.86\\
\end{tabular}
\end{ruledtabular}
\end{table}

Next, we analyze the xc torque along the out-of-plane direction of the cluster (the $z$ direction). Figure \ref{fig:torque} shows the exchange torque obtained with our MGGAx, compared with Slater and EXX-KLI.  Clearly, while the Slater potential provides decent magnetic moments, the exchange torque does not resemble that obtained by EXX-KLI, apart from regions close to the atoms.
As expected for our proposed MGGA, we obtain a nonzero exchange torque as a consequence of the fact that the curvature of the noncollinear exchange hole is not aligned with the magnetization direction. The alternation of positive and negative local torques leads to an overall zero torque, as required by the zero-torque theorem \cite{ZTT}. Overall, our functional agrees well with the  Slater/EXX-KLI torques in the regions around the atoms. In fact, our MGGA yields fewer torque features than Slater, in better agreement with EXX-KLI, especially in the interstitial region, except that the signs of the features are inverted. There are slight differences between the gKS and OEP
implementations of the MGGA, but these are small compared to the differences between MGGA and Slater/EXX-KLI.
Further away from the atoms, it is clear that the proposed MGGA does not capture all the details of the exchange torque, as expected for a functional based on a short-range expansion of the noncollinear exchange hole.

We note that compared to the spin-spiral wave noncollinear functional of Eich and Gross~\cite{PhysRevLett.111.156401}, which produces a six-fold symmetric exchange torque around all the Cr atoms, our functional yields a torque that depends on the local environment of the atoms, as obtained by the Slater potential, EXX-KLI, or EXX-OEP for Cr monolayers~\cite{PhysRevLett.98.196405}. The noncollinear GGA by Scalmani and Frisch \cite{scalmani2012new} produces an exchange torque depending on the local environment of the atoms, but their result for the torques displays a wrong number of positive-negative features around the atoms compared to those of Slater, EXX-KLI, and our MGGA. Overall, while our MGGAx results for Cr$_3$ leave some room for improvement, they do provide a realistic description of the exchange torques.
\Nrem{We also performed calculations for a Cr$_5$ cluster, see SM~\cite{supp}, with
similar conclusions as for Cr$_3$.}

\begin{figure*}
  \begin{center}
    \includegraphics[width=0.99\textwidth]{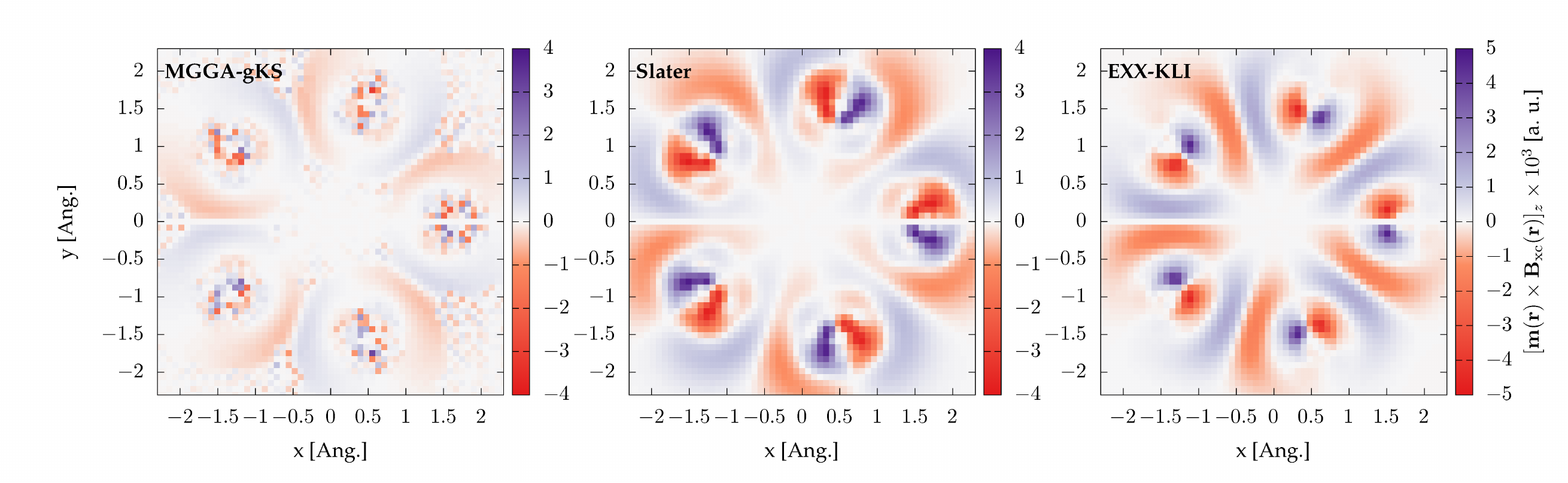}
    \caption{\label{fig:torque2} Out-of-plane component of the exchange torques for a Cr$_5$ cluster, calculated with MGGAx-gKS, Slater, and EXX-KLI.}
  \end{center}
\end{figure*}

We also performed calculations for a Cr$_5$ cluster, see Fig. \ref{fig:torque2}, using a distance of 3.7 Bohr between the Cr atoms.
Similar to Cr$_3$ (see main text), our exchange functional produces an exchange torque of the same order of magnitude and with similar features around the atoms as that obtained by Slater or EXX-KLI. As for Cr$_3$, the sign of the exchange torque is inverted in the interstitial region. Again, we attribute this to the short-range nature of our expansion, which has difficulties producing the correct sign of the $\mathbf{B}_{\mathrm{x}}$ away from the atomic centers.

\begin{figure}
  \begin{center}
    \includegraphics[width=\columnwidth]{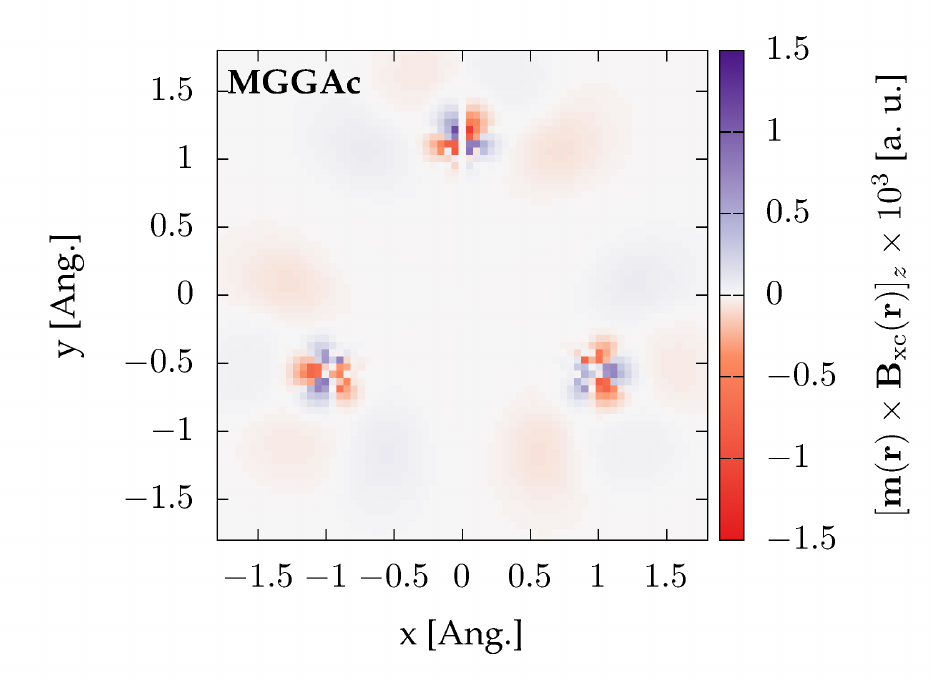}
    \caption{\label{fig:torque_c} Same as Fig.~\ref{fig:torque}, but now showing the correlation torque (at the MGGAc-gKS level).}
  \end{center}
\end{figure}

Figure \ref{fig:torque_c} presents the corresponding correlation torque for Cr$_3$. Similar conclusions are obtained as for the exchange torque (Fig.~\ref{fig:torque}), even if the magnitude of the correlation torque is smaller than for the exchange one. We note a sign change of the torque close to the atom between the exchange and correlation parts, indicating an intricate balance between exchange and correlation.

\section{Conclusions}

In conclusion, we have developed a semilocal xc functional for noncollinear SDFT
which belongs to the class of MGGAs, is numerically well behaved, and computationally much cheaper than nonlocal exchange functionals of comparable
quality (Slater and KLI). The correlation part of the functional provides the first insight on the shape and strength of the correlation torques.
Potential applications of this new functional range from \textit{ab initio} equilibrium studies of unconventional spin structures like skyrmions to a variety of dynamical phenomena in magnetic materials such as spin-wave dispersions
and the demagnetization in light-driven magnetic solids. From a more fundamental perspective, this work highlights the important and interesting
role of xc torques in SDFT, which should motivate further study.

\acknowledgments
N. T.-D. and A. R. acknowledge support from the European Research Council (ERC-2015-AdG694097), the Cluster of Excellence ``CUI: Advanced Imaging of Matter'' of the Deutsche Forschungsgemeinschaft (DFG) - EXC 2056 - project ID 390715994, Grupos Consolidados (IT1453-22) and SFB925. The Flatiron Institute is a division of the Simons Foundation.
C.A.U. was supported by DOE Grant No. DE-SC0019109.
We thank Giovanni Vignale and Stefano Pittalis for helpful comments.

\appendix

\section{Noncollinear exchange hole}
\label{app:exchange_hole}
The Slater potential for noncollinear spin systems, $\tens{v}^{\mathrm{S}}$, is defined by a Sylvester equation \cite{PhysRevB.98.035140}:
\begin{equation}
 \tens{v}^{\mathrm{S}}(\mathbf{r})\tens{n}(\mathbf{r}) + \tens{n}(\mathbf{r})\tens{v}^{\mathrm{S}}(\mathbf{r}) = 2 \int\frac{ d\mathbf{r'}}{|\mathbf{r} - \mathbf{r'}|} \tens{\gamma}(\mathbf{r},\mathbf{r'}) \tens{\gamma}(\mathbf{r'},\mathbf{r}) \,,
\end{equation}
where $\tens{\gamma}(\mathbf{r},\mathbf{r}')$ is the spin density matrix defined after Eq. (1) in the main text.
Similar to the collinear case, one can write the Slater potential as the Coulomb potential originating from an exchange hole, which here has the form of a tensor:
\begin{equation}
\tens{v}^{\mathrm{S}}(\mathbf{r}) = -\int \frac{ d\mathbf{r'}}{|\mathbf{r} - \mathbf{r'}|} \tens{\rho}^{\mathrm{x}}(\mathbf{r},\mathbf{r'})\,.
\end{equation}
From this we obtain a generalization of the exchange hole to the noncollinear case, $\tens{\rho}^{\mathrm{x}}$, defined by the equation
\begin{equation} \label{xhole_def}
\tens{n}(\mathbf{r}) \tens{\rho}^{\mathrm{x}}(\mathbf{r},\mathbf{r'}) + \tens{\rho}^{\mathrm{x}}(\mathbf{r},\mathbf{r'}) \tens{n}(\mathbf{r})
= 2  \tens{\gamma}(\mathbf{r},\mathbf{r'}) \tens{\gamma}(\mathbf{r'},\mathbf{r})\,.
\end{equation}
It is straightforward to verify that this definition leads to the required properties that one expects for the exchange hole. Moreover, this expression reduces to the usual one if we take the collinear limit.
From this definition, we can recast the exchange energy in terms of the noncollinear exchange hole:
\begin{equation}
 E_{\mathrm{x}} = - \frac{1}{4} \int \int \frac{d\mathbf{r}d\mathbf{r'}}{|\mathbf{r}-\mathbf{r'}|} \mathrm{Tr}\Big[ \tens{n}(\mathbf{r}) \tens{\rho}^{\mathrm{x}}(\mathbf{r},\mathbf{r'}) + \tens{\rho}^{\mathrm{x}}(\mathbf{r},\mathbf{r'}) \tens{n}(\mathbf{r})\Big] \,.
\end{equation}
However, while this expression is conceptually important, it does not allow for simple approximations as one needs to approximate both the diagonal and off-diagonal parts of the exchange hole, and we are not aware of existing models for the off-diagonal part.
This is why, in the main text [see Eq. (2)], we followed the alternative path of approximating an effective exchange hole.

\section{Hydrogenic model for the exchange hole}
\label{app:hydrogenic}
The hydrogenic model for the exchange hole proposed by Becke and Roussel \cite{PhysRevA.39.3761} reads as follows:
\begin{eqnarray}
 \rho_{\rm H}^{\rm x}(a,b;s) &=& \frac{a}{16\pi b s}\Big[(a|b-s| + 1) e^{-a|b-s|} \nonumber\\
&& - (a(|b+s|+1)e^{-a|b+s|}\Big]\,,
 \label{eq:model_pot}
\end{eqnarray}
where $a$ and $b$ are positive parameters of the model.
This model has two important properties: it is nonnegative, and it is normalized as
\begin{equation}
 4\pi \int ds \, s^2 \rho_{\rm H}^{\rm x}(a,b;s) = 1\,.
\end{equation}
As we showed in the main text, the effective exchange-hole matrix has the expected properties.

Imposing the on-top condition for the hydrogenic model, we find from the zeroth-order term of the Taylor expansion that
\begin{equation}
\frac{a^3}{8\pi}e^{-ab} = h_{\rm x}(\mathbf{r},\mathbf{r}) \,.
\end{equation}
The second-order term is defined by
\begin{equation}\label{8}
\frac{(a^2b-2a) }{6 b} = \frac{Q(\mathbf{r},\mathbf{r})}{h_{\rm x}(\mathbf{r},\mathbf{r})} \,,
\end{equation}
where $Q$ is the curvature of the effective exchange hole.
From the second-order term, and defining $x=ab$, one finds that $x$ satisfies the equation
\begin{equation}\label{9}
(x-2) = xe^{-2x/3} \frac{6 Q(\mathbf{r},\mathbf{r})}{4 \pi^{2/3}(h_{\rm x}(\mathbf{r},\mathbf{r}))^{5/3}} \,,
\end{equation}
and one also obtains
\begin{equation}\label{10}
 b^3 = \frac{x^3 e^{-x}}{8\pi h_{\rm x}(\mathbf{r},\mathbf{r})} \,.
\end{equation}
Equation (\ref{9}) can be put in a more familiar form:
\begin{equation}
\frac{xe^{-2x/3}}{(x-2)} = \frac{2}{3}\pi^{2/3}\frac{(h_{\rm x}(\mathbf{r},\mathbf{r}))^{5/3}}{Q(\mathbf{r},\mathbf{r})}\,.
\end{equation}
This is equation is numerically solved for $x$, which then yields the values for $b$ and $a$ via Eqs. (\ref{8}) and (\ref{10}).

\section{Noncollinear MGGA potential and nonlocal term}
\label{app:potential}
For the sake of generality, we consider here a generic form of a noncollinear MGGA depending on the spin density matrix, its gradient and Laplacian, as well as the kinetic energy density matrix,
\begin{equation}
 E_{\mathrm{xc}} = \int d\mathbf{r} e_{\mathrm{xc}}(\tens{n}(\mathbf{r}),\nabla \tens{n}(\mathbf{r}), \nabla^2 \tens{n}(\mathbf{r}), \tens{\tau}(\mathbf{r}))\,.
\end{equation}
Its functional derivative with respect to the density is
\begin{eqnarray}
v_{\mathrm{xc},\alpha\beta}(\mathbf{r})
&=&
\frac{\delta E_{\mathrm{xc}}}{ \delta n_{\beta\alpha}(\mathbf{r}) } \nonumber\\
&=& e_{\mathrm{xc},\beta\alpha}^1(\mathbf{r}) - \nabla e_{\mathrm{xc},\beta\alpha}^2(\mathbf{r}) + \nabla^2 e_{\mathrm{xc},\beta\alpha}^3(\mathbf{r}) \nonumber\\
&&+ \sum_{\gamma\delta}\int e_{\mathrm{xc},\gamma\delta}^4(\mathbf{r'}) \frac{\delta \tau_{\gamma\delta}(\mathbf{r'})}{\delta n_{\beta\alpha}(\mathbf{r})} d\mathbf{r'}\,,
\label{eq:potential}
\end{eqnarray}
where we defined
\begin{eqnarray}
e_{\mathrm{xc},\nu\mu}^1(\mathbf{r}) &=& \frac{\partial e_{\mathrm{xc}}(\tens{n},\nabla \tens{n}, \nabla^2 \tens{n}, \tens{\tau})}{\partial n_{\nu\mu}(\mathbf{r})} \\[2mm]
e_{\mathrm{xc},\nu\mu}^2(\mathbf{r}) &=& \frac{\partial e_{\mathrm{xc}}(\tens{n},\nabla \tens{n}, \nabla^2 \tens{n}, \tens{\tau})}{\partial \nabla n_{\nu\mu}(\mathbf{r})} \\[2mm] e_{\mathrm{xc},\nu\mu}^3(\mathbf{r}) &=& \frac{\partial e_{\mathrm{xc}}(\tens{n},\nabla \tens{n}, \nabla^2 \tens{n}, \tens{\tau})}{\partial \nabla^2 n_{\nu\mu}(\mathbf{r})}\\[2mm] e_{\mathrm{xc},\nu\mu}^4(\mathbf{r}) &=& \frac{\partial e_{\mathrm{xc}}(\tens{n},\nabla \tens{n}, \nabla^2 \tens{n}, \tens{\tau})}{\partial \tau_{\nu\mu}(\mathbf{r})} \:.
\end{eqnarray}
One important aspect of noncollinear MGGAs lies in the evaluation of the contribution to the xc potential originating from the functional derivative of the kinetic energy density with respect to the spin density matrix, $v^\tau$, which is given by the last term of Eq. (\ref{eq:potential}):
\begin{equation}
 v^\tau_{\alpha\beta}(\mathbf{r}) = \sum_{\gamma\delta}\int e_{\mathrm{xc},\nu\mu}^4(\mathbf{r}) \frac{\delta \tau_{\gamma\delta}(\mathbf{r'})}{\delta n_{\beta\alpha}(\mathbf{r})} d\mathbf{r'}\,,
\end{equation}
which we do not know how to compute directly. Here $\alpha, \beta, \gamma, \delta$ refer to spin indices. In order to evaluate the application of the MGGA potential
to a given orbital, in the framework of generalized Kohn-Sham equations, we need a generalization of the approach used in the collinear case.
In the collinear case, we have the relation
\begin{equation}
 \frac{\delta E_{\mathrm{xc}}}{\delta \psi^*_{i}(\mathbf{r})} = \frac{\delta E_{\mathrm{xc}}}{\delta n(\mathbf{r})} \psi_{i}(\mathbf{r})\,,
\end{equation}
which is based on the fact that $\frac{\delta n(\mathbf{r})}{\delta \psi^*_{i}(\mathbf{r'})} = \psi_{i}(\mathbf{r}) \delta(\mathbf{r'}-\mathbf{r})$.

In the noncollinear case, the spin density matrix is given by $n_{\alpha\beta}(\mathbf{r}) = \sum_i \psi_{i\alpha}(\mathbf{r})\psi^*_{i\beta}(\mathbf{r})$, where $\psi_{i}(\mathbf{r})$ is a Pauli spinor describing the $i$-th orbital of the electronic system. From the definition of the spin density matrix it then directly follows that
\begin{equation}
 \frac{\delta n_{\alpha\beta}(\mathbf{r})}{\delta \psi^*_{i\sigma}(\mathbf{r'})} = \psi_{i\alpha}(\mathbf{r})\delta_{\beta,\sigma} \delta(\mathbf{r'}-\mathbf{r})\,.
\end{equation}

In the case of a pure functional of the density, using the chain rule, we obtain that
\begin{eqnarray}
 \frac{\delta E_{\mathrm{xc}}}{\delta \psi^*_{i\sigma}(\mathbf{r})} &=& \sum_{\alpha\beta} \int d\mathbf{r'} \frac{\delta E_{\mathrm{xc}}}{\delta n_{\alpha\beta}(\mathbf{r'})}  \frac{\delta n_{\alpha\beta}(\mathbf{r})}{\delta \psi^*_{i\sigma}(\mathbf{r'})} \nonumber\\
 &=& \sum_{\alpha}  \frac{\delta E_{\mathrm{xc}}}{\delta n_{\alpha\sigma}(\mathbf{r})} \psi_{i\alpha}(\mathbf{r})
 =
 \sum_{\alpha} (v_{\mathrm{xc},\sigma\alpha} \psi_{i\alpha})(\mathbf{r}) \nonumber
 \,.
\end{eqnarray}
We recognize here the application of the xc potential to a Pauli spinor, for which we take the component $\sigma$. It is clear that the connection between this quantity and the potential is less obvious than in the collinear case.

Let us now consider the case of a pure MGGA depending only on the kinetic energy density.
Using the definition of the noncollinear kinetic energy (without the factor $1/2$), $\tau_{\alpha\beta} = \sum_i (\nabla \psi_{i\alpha})\cdot(\nabla\psi_{i\beta})^*$, we get
\begin{equation}
 \frac{\delta \tau_{\alpha\beta}}{\delta \psi^*_{i\sigma}(\mathbf{r'})} = (\nabla \psi_{i\alpha}) \delta_{\beta,\sigma} \nabla \delta(\mathbf{r}-\mathbf{r'}) .
\end{equation}
We therefore have in this case
\begin{eqnarray}
 \frac{\delta E_{\rm xc}[\tau]}{\delta \psi^*_{i\sigma}(\mathbf{r})}
 &=& \sum_{\alpha\beta} \int d\mathbf{r'}  \frac{\delta E_{\rm xc}[\tau]}{\delta \tau_{\alpha\beta}(\mathbf{r'})}  \frac{\delta \tau_{\alpha\beta}(\mathbf{r'})}{\delta \psi^*_{i\sigma}(\mathbf{r})} \nonumber\\
 &=&  \sum_{\alpha} \int d\mathbf{r'}  \frac{\delta E_{\rm xc}[\tau]}{\delta \tau_{\alpha\sigma}(\mathbf{r'})}  (\nabla \psi_{i\alpha}(\mathbf{r'})) \nabla \delta(\mathbf{r}-\mathbf{r'}) \nonumber\\
 &=& -\sum_{\alpha}  \nabla \cdot \Big( \frac{\delta E_{\rm xc}[\tau]}{\delta \tau_{\alpha\sigma}(\mathbf{r})} \nabla \psi_{i\alpha}(\mathbf{r'}) \Big)\,.
\end{eqnarray}
This leads to the result that
\begin{eqnarray}
  \frac{\delta E_{\rm xc}[\tau]}{\delta n_{\alpha\sigma}(\mathbf{r})} \psi_{i\alpha}(\mathbf{r}) &=& -\nabla\cdot \Big( \frac{\delta E_{\rm xc}[\tau]}{\delta \tau_{\alpha\sigma}(\mathbf{r})}  \nabla \psi_{i\alpha}(\mathbf{r}) \Big)\nonumber\,,
\end{eqnarray}
which is the noncollinear analog of the collinear relation, see for instance Ref.~\cite{doi:10.1063/1.4811270}.

From this it is straightforward to show that for a most general form of a noncollinear MGGA, we have
\begin{eqnarray}
 \frac{\delta E_{\mathrm{x}}}{ \psi^*_{i,\sigma}} &=& \sum_\alpha\Big(e_{\rm x,\alpha\sigma}^1(\mathbf{r}) - \nabla e_{\rm x,\alpha\sigma}^2(\mathbf{r}) + \nabla^2 e_{\rm x,\alpha\sigma}^3(\mathbf{r})\Big)\psi_{i,\sigma} \nonumber\\
 &&-\sum_\alpha \nabla\cdot \Big( e_{\rm x,\alpha\sigma}^4(\mathbf{r})  \nabla \psi_{i\alpha}(\mathbf{r}) \Big)\,,
\end{eqnarray}
where the first part is defining the local xc potential and the second part corresponds to the nonlocal term originating from the dependence on the kinetic energy density.

\section{One-electron limit of the exchange-hole curvature}
\label{app:one_electron}
We now consider the one-electron limit of the exchange-hole curvature,
\begin{equation}
 Q = \frac{1}{6} \Big[\nabla^2n -2\gamma D\Big]\,,
 \nonumber
\end{equation}
with
\begin{equation}
D = \bar{\tau}-\frac{\nabla n\cdot\nabla n}{4n} = \bar{\tau}-\tau^W
\end{equation}
and
\begin{eqnarray}
n\bar{\tau}
&=&
\frac{1}{2}\sum_\sigma \bigg[2n_{\sigma\sigma}\tau_{\sigma\sigma} +n_{\sigma-\sigma}\tau_{-\sigma\sigma}
+ \tau_{\sigma-\sigma}n_{-\sigma\sigma} \bigg]\nonumber\\
&-&
\sum_\sigma \bigg[|\mathbf{j}_{\sigma\sigma}|^2 +|\mathbf{j}_{\sigma-\sigma}|^2 \bigg]
+
\frac{1}{4}\sum_\sigma\bigg[
 2n_{\sigma\sigma}\nabla^2n_{-\sigma-\sigma} \nonumber\\[2mm]
&&
{}-(n_{\sigma-\sigma}\nabla^2n_{-\sigma\sigma}
 + \nabla^2n_{\sigma-\sigma}n_{-\sigma\sigma})  \nonumber\\[2mm]
&&
{}-|\nabla n_{\sigma-\sigma}|^2
 +\nabla n_{\sigma\sigma}\cdot\nabla n_{-\sigma-\sigma}\bigg] \,.
\end{eqnarray}
Let us now look at the one-electron limit of the noncollinear kinetic energy density:
\begin{equation}
 \tau_{\alpha\beta} \to (\nabla \psi_\alpha)\cdot(\nabla\psi_\beta^*)  \:.
\end{equation}
Using the fact that
\begin{subequations}
\begin{align}
 \nabla n_{\alpha\beta} \to (\nabla \psi_\alpha) \psi_\beta^* + \psi_\alpha \nabla\psi_\beta^* \\
 \mathbf{j}_{\alpha\beta} \to \frac{1}{2i}[(\nabla\psi_\alpha)\psi_\beta^* - \psi_\alpha\nabla\psi_\beta^*]
\end{align}
\end{subequations}
we get
\begin{subequations}
\begin{align}
 \frac{1}{2}(\nabla n_{\alpha\beta} -2i\mathbf{j}_{\alpha\beta}) \to \psi_\alpha \nabla\psi_\beta^* \\
 \frac{1}{2}(\nabla n_{\alpha\beta} +2i\mathbf{j}_{\alpha\beta}) \to  (\nabla \psi_\alpha) \psi_\beta^* \:.
\end{align}
\end{subequations}
From these expressions it is clear that
\begin{equation}
 n_{\sigma\sigma}\tau_{\sigma\sigma} -|\mathbf{j}_{\sigma\sigma}|^2  = \frac{1}{4}|\nabla n_{\sigma\sigma}|^2\,,
 \label{eq:collinear_one_electron}
\end{equation}
which is the known result for collinear case.
Similarly, one finds that
\begin{eqnarray}
\lefteqn{\hspace{-1.5cm} n_{\sigma-\sigma}(\nabla^2n_{-\sigma\sigma}-2\tau_{-\sigma\sigma})+ (\nabla^2n_{-\sigma\sigma} -2\tau_{\sigma-\sigma})n_{-\sigma\sigma}} \nonumber\\
  &\to& n_{\sigma\sigma}(\nabla^2n_{-\sigma-\sigma}-2\tau_{-\sigma-\sigma})\nonumber\\
  &&{} + n_{-\sigma-\sigma}(\nabla^2n_{\sigma\sigma}-2\tau_{\sigma\sigma})\:.
\end{eqnarray}


Combining these expressions, we obtain the following result for the one-electron limit:
\begin{eqnarray}
 nD \to -\sum_\sigma \Big[ |\mathbf{j}_{\sigma-\sigma}|^2
 +\frac{1}{4}|\nabla n_{\sigma-\sigma}|^2 +\tau_{\sigma\sigma}n_{-\sigma-\sigma} \Big] = 0\,. \nonumber
\end{eqnarray}
Therefore $D$ vanishes in the one-electron limit, which shows that in this limit we have $\bar{\tau}=\tau^W$.

\vspace{1cm}

\section{Uniform gas limit}

In the original Becke-Roussel paper~\cite{PhysRevA.39.3761}, the value of $\gamma=0.8$ was introduced to improve the agreement of the exchange hole with respect to the uniform electron gas exchange hole. As our functional recovers the collinear spin limit, it is already justified to use this value. Moreover, we can show that the structure of the noncollinear exchange hole in the uniform electron gas limit is similar to the one of the unpolarized or collinear cases.

For this, we consider the limit of the uniform electron gas. In this limit, the kinetic energy density
limit $\tau^{\mathrm{unif}}_{\sigma\sigma'}$ becomes
\begin{equation}
 \tau^{\mathrm{unif}}_{\sigma\sigma'} = \frac{3}{5}k_\mathrm{F}^2 n_{\sigma\sigma'} \,,
\end{equation}
where $k_\mathrm{F}$ is the Fermi vector of our uniform electron gas, given by
\begin{equation}
k_\mathrm{F} = (6\pi^2 n)^{1/3} \:.
\end{equation}
By setting $\nabla n_{\sigma\sigma'} =0$, and $\nabla^2 n_{\sigma\sigma'} = 0$, we arrive at
\begin{eqnarray}
\bar{\tau}^{\mathrm{unif}} &=& \frac{3}{5n} k_\mathrm{F}^2 \left( n_{\uparrow\uparrow}^2 + n_{\downarrow\downarrow}^2 + 2|n_{\uparrow\downarrow}|^2   \right)
\nonumber\\
&=& \frac{3}{5} k_\mathrm{F}^2 \left( \frac{n}{2} + \frac{|\mathbf{m}|^2}{2n} \right),
\end{eqnarray}
where we recognize the on-top value of the exchange hole ($n/2 + |\mathbf{m}|^2/2n$) of the last term. Therefore, it is clear that the exchange hole
curvature behaves similarly to the unpolarized case, as long as we replace the on-top curvature of the noncollinear system by the local density of the unpolarized system. This further justify the use of $\gamma=0.8$ in our functional.

\begin{widetext}
\section{Explicit solution of the KLI equation in the noncollinear case}
\label{app:KLI}
The noncollinear version of the KLI potential is defined as \cite{PhysRevB.98.035140}
\begin{eqnarray}
\lefteqn{
 [\tens{n}(\mathbf{r})\tens{v}^{\rm KLI}(\mathbf{r}) + \tens{v}^{\rm KLI}(\mathbf{r})\tens{n}(\mathbf{r})]_{\nu\mu} = \sum_i \Big( \frac{\delta E_{\rm xc}}{\delta \psi^*_{i\nu}(\mathbf{r})}\psi^*_{i\mu}(\mathbf{r}) + \frac{\delta E_{\rm xc}}{\delta \psi_{i\mu}(\mathbf{r})}\psi_{i\nu}(\mathbf{r})\Big)}
 \nonumber\\
&+& \sum_i \psi_{i\nu}(\mathbf{r})\psi^*_{i\mu}(\mathbf{r})\int d\mathbf{r'} \Big(2\sum_{\alpha\beta}v^{\rm KLI}_{\alpha\beta}(\mathbf{r'})\psi_{i\alpha}^*(\mathbf{r'})\psi_{i\beta}(\mathbf{r'}) - \sum_{\alpha}\frac{\delta E_{\rm xc}}{\delta \psi^*_{i\alpha}(\mathbf{r'})}\psi^*_{i\alpha}(\mathbf{r'})  - \sum_{\alpha}\frac{\delta E_{\rm xc}}{\delta \psi_{i\alpha}(\mathbf{r'})}\psi_{i\alpha}(\mathbf{r'})\Big)\,.
\end{eqnarray}
If only the first part on the right-hand side is taken into account then the KLI potential reduces to the Slater potential, $\tens{v}^{\rm KLI} \to \tens{v}^{\rm S}$.
We now seek a set of orbital-dependent constants $\{C_{i}\}$, such that
\begin{equation} \label{35}
  [\tens{n}(\mathbf{r})\tens{v}^{\rm KLI}(\mathbf{r}) + \tens{v}^{\rm KLI}(\mathbf{r})\tens{n}(\mathbf{r})]_{\nu\mu} = [\tens{n}(\mathbf{r})\tens{v}^{\rm S}(\mathbf{r}) + \tens{v}^{\rm S}(\mathbf{r})\tens{n}(\mathbf{r})]_{\nu\mu} + \sum_i \rho_{i\nu\mu}(\mathbf{r}) C_{i}\,,
\end{equation}
where we defined $\rho_{i\nu\mu}(\mathbf{r}) = \psi_{i\nu}(\mathbf{r})\psi^*_{i\mu}(\mathbf{r})$.
It is clear that
\begin{equation}
  C_{i} = \int d\mathbf{r'} \Big(2\sum_{\alpha\beta}v^{\rm KLI}_{\alpha\beta}(\mathbf{r'})\rho_{i\beta\alpha}(\mathbf{r'})  - \sum_{\alpha}\frac{\delta E_{\rm xc}}{\delta \psi^*_{i\alpha}(\mathbf{r'})}\psi^*_{i\alpha}(\mathbf{r'})  - \sum_{\alpha}\frac{\delta E_{\rm xc}}{\delta \psi_{i\alpha}(\mathbf{r'})}\psi_{i\alpha}(\mathbf{r'})\Big)
   = \bar{V}_{i} - \bar{v}^{\rm S}_{i}\,,
\end{equation}
where $\bar{V}_{i,\alpha\beta}$ needs to be determined.
\end{widetext}

Assuming that the density matrix is not singular, this can be solved as \cite{Krieger1992}
\begin{equation}
  v^{\rm KLI}_{\nu\mu}(\mathbf{r}) = v^{\rm S}_{\nu\mu}(\mathbf{r}) + \sum_i [\tens{\mathcal{N}}^{-1}(\mathbf{r}) \mbox{\bf vec} \tens{\rho_i}(\mathbf{r})]_{\nu\mu} C_{i}\,,
  \label{eq:kli_nonsingular}
\end{equation}
where $\mbox{\bf vec} \tens{\rho_i}$ rearranges the elements of the $2\times 2$ matrix $\tens{\rho_i}$ into a single column vector with four components,
$\tens{\mathcal{N}}$ is a $4\times 4$ matrix given by \cite{PhysRevB.98.035140}
\begin{equation}
\tens{\mathcal{N}} = \tens{I} \otimes \tens{n} + \tens{n}^T \otimes \tens{I}  \:,
\end{equation}
and $\tens{I}$ is the $2\times 2$ unit matrix.
We now multiply both sides of Eq. (\ref{eq:kli_nonsingular}) by $2\rho_{j\mu\nu}(\mathbf{r})$, sum over the spin indices, and integrate over $\mathbf{r}$.
This gives
\begin{eqnarray}
 \bar{V}_{j} &=& 2 \sum_{\nu\mu} \int d\mathbf{r} \rho_{j\mu\nu}(\mathbf{r}) v^{\rm S}_{\nu\mu}(\mathbf{r}) \\
  &+& 2 \sum_i  \sum_{\nu\mu} \int d\mathbf{r} \rho_{j\mu\nu}(\mathbf{r}) [\tens{\mathcal{N}}^{-1}(\mathbf{r})  \mbox{\bf vec}\tens{\rho_i}(\mathbf{r})]_{\nu\mu} (\bar{V}_{i} - \bar{v}^{\rm S}_{i}) \:.
  \nonumber
\end{eqnarray}
We subtract  $\bar{v}^{\rm S}_{j}$ on both sides and obtain after some rearrangement
\begin{align}
 \sum_i \Big[ \delta_{ij} - 2 \sum_{\nu\mu} \int d\mathbf{r} \rho_{j\mu\nu}(\mathbf{r}) [\tens{\mathcal{N}}^{-1}(\mathbf{r}) \mbox{\bf vec} \tens{\rho_i}(\mathbf{r})]_{\nu\mu} \Big] C_{i}  \nonumber\\
 =  2 \sum_{\nu\mu} \int d\mathbf{r} \rho_{j\mu\nu}(\mathbf{r}) v^{\rm S}_{\nu\mu}(\mathbf{r}) -  \bar{v}^{\rm S}_{j} \:.
\end{align}
We recognize a linear equation that can be solved straightforwardly, similar to the collinear case \cite{Krieger1992}.

Now let us consider the case for which the density matrix is singular.
In this case, the Sylvester equation cannot be solved directly.
To treat this problem, we perform a local rotation of the equation in the magnetization frame, defined by the rotation matrix $\tens{R}$. In this frame, the density matrix is zero expect for one diagonal element and the corresponding matrix is denoted $\tens{D} = \tens{R}^\dag\tens{n}\tens{R}$. The potentials in this frame are labeled by the superscript ``loc''.
Equation (\ref{35}) then becomes
\begin{equation}
 [\tens{D}\tens{v}^{\rm KLI,loc}+ \tens{v}^{\rm KLI,loc}\tens{D}] = [\tens{D}\tens{v}^{\rm S,loc} + \tens{v}^{\rm S,loc}\tens{D}] + \sum_i \tens{\rho_i}^{\rm loc} C_{i}
\end{equation}
In the case of $\tens{D} =
\begin{pmatrix}
 n_{\uparrow} & 0 \\
 0 & 0
\end{pmatrix}$, we obtain
\begin{eqnarray}
n_{\uparrow} \begin{pmatrix}
  2 v^{\rm KLI,loc}_{\uparrow\uparrow} &  v^{\rm KLI,loc}_{\uparrow\downarrow}  \\
  v^{\rm KLI,loc}_{\downarrow\uparrow}  & 0
\end{pmatrix}
&=&
n_{\uparrow}\begin{pmatrix}
  2  v^{\rm S,loc}_{\uparrow\uparrow} &  v^{\rm S,loc}_{\uparrow\downarrow}  \\
   v^{\rm S,loc}_{\downarrow\uparrow}  & 0
\end{pmatrix} \nonumber\\
&+&
\sum_{i}  \tens{\rho_i}^{\rm loc} C_{i} \:.
\end{eqnarray}
This gives us two equations to solve:
\begin{eqnarray}\label{eq:kli_singular1}
 v^{\rm KLI,loc}_{\uparrow\uparrow}(\mathbf{r}) = v^{\rm S,loc}_{\uparrow\uparrow}(\mathbf{r}) + \sum_{i} \frac{\rho_{i\uparrow\uparrow}^{\rm loc}(\mathbf{r}) C_{i}}{2n_{\uparrow}(\mathbf{r})}\\
 v^{\rm KLI,loc}_{\uparrow\downarrow}(\mathbf{r}) = v^{\rm S,loc}_{\uparrow\downarrow}(\mathbf{r}) + \sum_{i} \frac{\rho_{i\uparrow\downarrow}^{\rm loc}(\mathbf{r}) C_{i}}{n_{\uparrow}(\mathbf{r})}
 \label{eq:kli_singular2}
\end{eqnarray}
and a similar equation for $v^{\rm KLI,loc}_{\downarrow\uparrow}$.
The last component, $v^{\rm KLI,loc}_{\downarrow\downarrow}$ is set to zero.
This equation is solved similar to the collinear case, and once the solution is obtained, we rotate the potential back to the original frame.

Combining these two approaches, we can write a general explicit solution for orbital-dependent constants needed for the KLI potential of the form
\begin{align}
 \sum_i \Big[ \delta_{ij} - 2 \sum_{\nu\mu} \int d\mathbf{r} \rho_{j\mu\nu}(\mathbf{r}) M_{i\nu\mu}({\bf r}) \Big] C_{i}  \nonumber\\
 =  2 \sum_{\nu\mu} \int d\mathbf{r} \rho_{j\mu\nu}(\mathbf{r}) v^{\rm S}_{\nu\mu}(\mathbf{r}) -  \bar{v}^{\rm S}_{j}\,,
\end{align}
where $M_{i\nu\mu}(\mathbf{r}) = [\tens{\mathcal{N}}^{-1}(\mathbf{r}) \mbox{\bf vec}\tens{\rho_i}(\mathbf{r})]_{\nu\mu}$ for the nonsingular points, and
\begin{displaymath}
\tens{M_{i}}(\mathbf{r}) = \tens{R}(\mathbf{r}) \Big( \frac{(1-\frac{1}{2}\tens{I})}{n_{\sigma}(\mathbf{r})}\tens{\rho}_{i}^{\rm loc}(\mathbf{r})\Big) \tens{R}^\dag(\mathbf{r})
\end{displaymath}
for the singular points.
The solution for the KLI potential is then constructed using either Eqs. (\ref{eq:kli_singular1}) and (\ref{eq:kli_singular2}) or Eq. (\ref{eq:kli_nonsingular}), respectively,
depending on whether the spin density matrix at a given point is singular or not.


%

\end{document}